\documentclass[prl,amsmath,amssymb,twocolumn]{revtex4-1}
\usepackage{graphicx}
\usepackage{epsfig}
\usepackage{dcolumn}
\usepackage{bm}
\usepackage{rotating}
\usepackage{multirow}
\usepackage[table]{xcolor}
\usepackage[english]{babel}

\begin{document}
\widetext 

\title{Thermally-Enhanced Fr\"ohlich Coupling in SnSe}

\author{Fabio Caruso}
\author{Maria Troppenz}
\author{Santiago Rigamonti}
\author{Claudia Draxl}
\affiliation{Institut f\"ur Physik and IRIS Adlershof, Humboldt-Universit\"at zu Berlin, Berlin, Germany} 
\date{\today}

\date{\today}
\begin{abstract}
To gain insight into the peculiar temperature dependence of
the thermoelectric material SnSe, we employ many-body 
perturbation theory and explore the influence of the electron-phonon 
interaction on its electronic and transport properties.
We show that a lattice dynamics characterized 
by soft highly-polar phonons induces a large thermal
enhancement of the {Fr\"ohlich} interaction. 
{We account for these phenomena in ab-initio calculations 
of the photoemission spectrum and electrical conductivity at finite 
temperature, unraveling the mechanisms behind recent experimental data.
Our results reveal a complex interplay between lattice thermal 
expansion and Fr\"ohlich coupling, providing a new rationale for 
the {\it in-silico} prediction of transport coefficients of
high-performance thermoelectrics.}
\end{abstract}

\maketitle

The discovery of the record-breaking thermoelectric properties of 
SnSe has laid a new milestone in the quest for high-efficiency 
thermoelectric materials \cite{Zhao2014}. 
SnSe combines large carrier conductivity $\sigma$ and Seebeck coefficient $S$, 
with highly anharmonic lattice dynamics \cite{Li2015,Zhao/review/2016,ph-anarm/ph3py/2016}.
Anharmonic effects limit the lattice thermal conductivity $\kappa$ via phonon-phonon scattering, thus 
contributing to a record-high figure of merit $ZT=(S^2\sigma/\kappa)T\sim 2.6$, which may 
be even further improved through doping and alloying \cite{doping/science/2015,JACS2018/Sn-vac,alkali/dop/JACS/2016,alloying/APL/2017}.

As the operational conditions of thermoelectric devices typically involve
large temperatures, a quantum-mechanical description of 
the electronic and lattice properties of SnSe across the temperature domain
of its thermodynamical stability is key to unravel the microscopic 
origin of this outstanding thermoelectric performance.
{Recent experimental investigations have unveiled a pervasive 
influence of temperature on the electronic and transport properties of SnSe. 
Angle-resolved photoemission spectroscopy experiments, 
for instance, have reported (i) a pronounced dependence 
of the peak linewidth on the sample temperature \cite{ARPES/meff/2017},
(ii) the emergence at low temperature of a gap between the first two 
occupied bands at the $Z$ point, \cite{ARPES/Nagayama2018,ARPES/Tayari/2018,ARPES/Wang2018,ARPES/band/conv/PRB2017}, 
and (iii) a non-monotonic effective-mass renormalization as a function of temperature \cite{Kanatzidis/2018}. 
Density-functional theory calculations fail at unraveling the origin of these phenomena. 
Additionally, theoretical predictions based on the Boltzmann formalism 
indicate an increase of the electrical conductivity 
with temperature -- arising from the thermal excitation of 
carriers across the Fermi surface -- which is in stark contrast 
with experimental observations, 
where a pronounced reduction of charge conduction with increasing temperature 
has been observed \cite{doping/science/2015}.

These findings seem to suggest a strong interplay between
electronic and ionic degrees of freedom.
While recent theoretical works have thus far provided a comprehensive investigation,  based on
first-principles calculations, of quasiparticle 
bands \cite{Kioupakis/QP/2015}, defect formation energies \cite{Huang/defects/2017},
crystal-lattice dynamics \cite{th/phon/2015}, lattice anharmonicities \cite{ph-anarm/ph3py/2016}, 
electron-phonon interaction \cite{Ma2018}, 
and transport properties \cite{band/transport/2015,Kioupakis/QP/2015,Verstraete/2016}, 
unraveling the origin of the peculiar temperature-dependent 
properties of this high-performance thermoelectric continues to 
represent a major challenge.}

In this work, we tackle the temperature-induced 
renormalization of the electronic and transport properties 
of SnSe by combining, within a first-principles framework, 
the many-body theory of the electron-phonon interaction 
in the Fan-Migdal approach and the Boltzmann transport 
formalism beyond the constant relaxation time, 
including effects of the lattice expansion. 
We find a strong enhancement of the electron-phonon 
coupling strength with temperature which stems from the thermal 
excitation of soft polar modes and manifests itself through 
band-structure renormalization effects, an increase of 
electron linewidths by a factor of five between 0 and 600~K, 
and a highly-anisotropic renormalization of the hole effective mass.
Our first-principles calculations of the electrical conductivity are
in excellent agreement with experiments, and demonstrate the importance 
of accounting simultaneously for the temperature-dependence
of the scattering time due to electron-phonon interaction
and the thermal expansion of the crystal lattice.
Overall, our results unveil a complex interplay between 
lattice anharmonicities and electron-phonon coupling which 
underpins the microscopic origin of the temperature-dependent 
properties of SnSe, paving the way towards the study of transport 
phenomena in thermoelectric materials. 

\begin{figure}[t] 
     \includegraphics[width=0.45\textwidth]{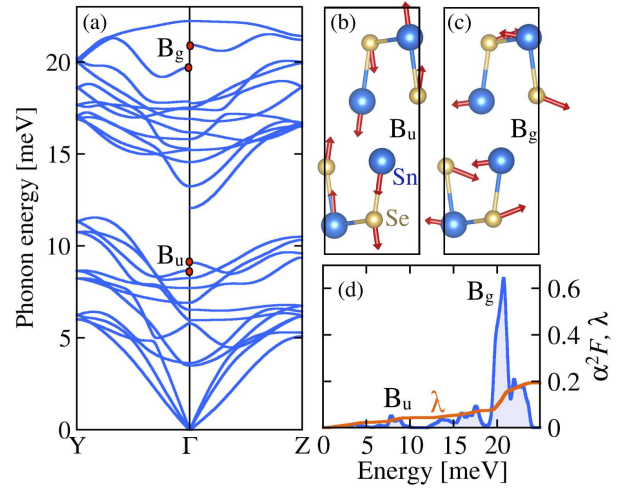} 
     \caption{(a) Phonon dispersion as obtained from DFPT. 
     (b)-(c) Eigenvectors of the polar phonons $B_u$ and $B_g$, respectively, marked by 
     red dots in (a). (d) Eliashberg function and {cumulative} electron-phonon 
     coupling strength $\lambda$ for doped SnSe. 
     }\label{fig:phon}
  \end{figure}

SnSe crystallizes in a layered orthorhombic 
structure of the {\it Pnma} space group \cite{Mariano1967}.
Its phonon dispersion, as obtained from density-functional 
perturbation theory (DFPT), is
illustrated in Fig.~\ref{fig:phon}~(a) for momenta along 
the high-symmetry line  Y-$\Gamma$-Z in the Brillouin zone. 
{Phonons have been computed for the 
relaxed crystal structure at zero temperature.}
SnSe is characterized by soft polar phonon modes with 
energies smaller than 25~meV (202~cm$^{-1}$). 
The polar character of the optical phonons 
manifests itself through the discontinuity 
in their energy at $\Gamma$ and 
is quantified by large average Born effective 
charges (-3.8 for Sn and 3.8 for Se, 
in good agreement with previous calculations \cite{Li2015}). 
In Fig.~\ref{fig:phon}~(b) and (c) arrows illustrate  
the atomic displacements corresponding to the $B_u$ 
and $B_g$ modes, marked by red dots in Fig.~\ref{fig:phon}~(a), 
which exemplify the lattice dynamics induced by polar phonons in SnSe. 
A comprehensive characterization of the optical  
phonon modes is provided in the Supplemental Material \cite{sup}. 

The pronounced polar character of SnSe suggests that 
the effects of electron-phonon interaction may be important. 
Additionally, since the energy scales of the optical 
modes are comparable to the room-temperature 
thermal energy $k_{\rm B}$T$\simeq 25$~meV,
the population of low-energy phonons, 
governed by the Bose-Einstein distribution 
$n_{{\bf q}\nu} = [ {\rm exp}({\hbar\omega_{{\bf q}\nu}}/{k_{\rm B}{\rm T}}) -1]^{-1}$,
is subject to a strong temperature dependence, which 
may lead to a significant thermal enhancement of the 
electron-phonon interaction. 
To investigate these aspects on a quantitative ground, 
we perform first-principles calculations of the 
electron-phonon interaction by evaluating the 
electron self-energy in the Fan-Migdal approximation as:
\begin{align}\label{eq:sigma}
  &\Sigma_{n{\bf k}} (\omega) = \int\!\frac{d{\bf q}}{\Omega_{\rm BZ}}\sum_{m\nu}
     |g_{mn\nu}({\bf k},{\bf q})|^2 \\
& \times  \left[ \frac { n_{{\bf q}\nu} + f_{m{\bf k+q}} }
  {\hbar\omega - \tilde\varepsilon_{m{\bf k+q}} + \hbar\omega_{{\bf q}\nu} }
  + \frac { n_{{\bf q}\nu} + 1 - f_{m{\bf k+q}} }
  {\hbar\omega - \tilde\varepsilon_{m{\bf k+q}} - \hbar\omega_{{\bf q}\nu} } \right].\nonumber
\end{align}
Here $ n_{{\bf q}\nu}$ and $f_{n{\bf k}}$ are Bose-Einstein and Fermi-Dirac 
occupation factors for phonons and electrons, respectively.  
$\tilde\varepsilon_{n{\bf k}} = \varepsilon_{n{\bf k}} + i\eta$,
where $\varepsilon_{n{\bf k}}$ are Kohn-Sham (KS) single particle 
energies, $\eta$ is a positive infinitesimal,  and
$\omega_{{\bf q}\nu}$ are phonon frequencies.
$g_{mn\nu}({\bf k},{\bf q}) = \left\langle \psi_{m{\bf k+q}} | 
\Delta_{{\bf q}\nu} V_{\rm KS}({\bf r}) | \psi_{n{\bf k}} \right\rangle$ 
are the electron-phonon coupling matrix elements, where $\Delta_{{\bf q}\nu} V_{\rm KS}$ 
denotes the change of the effective KS potential upon a 
phonon displacement with momentum ${\bf q}$ and index $\nu$, and $\psi_{n{\bf k}}$ are KS orbitals.
The polar character of the electron-phonon matrix elements $g_{mn\nu}({\bf k},{\bf q}) $ 
due to the Fr\"ohlich interaction is accounted for following the  
first-principles method of Ref.~\cite{Verdi2017}.
Equation~\eqref{eq:sigma} has been evaluated within the {\tt EPW} code \cite{Ponce2016a}. 
All computational details are provided in the Supplemental Material \cite{sup}. 

\begin{figure*}[t] 
\includegraphics[width=0.98\textwidth]{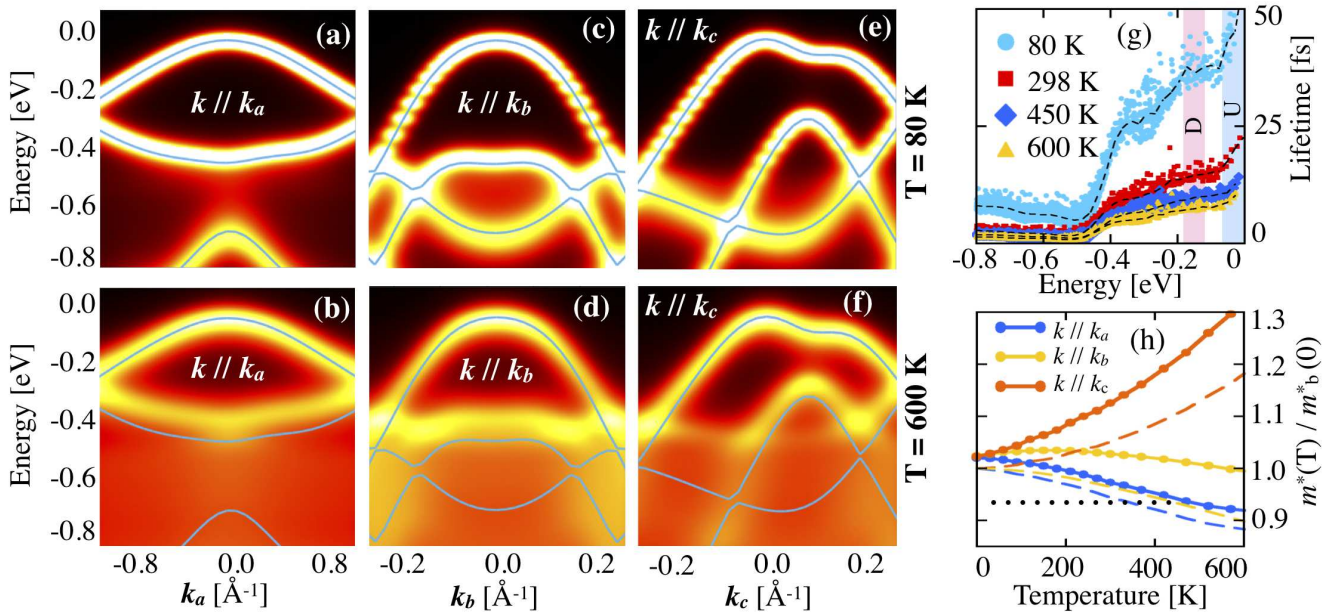} 
\caption{Temperature dependence of the angle-resolved spectral function for the valence bands for 
crystal momenta along the ${\bf k}_x$ [(a)-(b)], ${\bf k}_y$ [(c)-(d)], and  ${\bf k}_z$ [(e)-(f)] directions in
the Brillouin zone. The DFT bands (light blue) are superimposed for comparison. 
{Discrete features in panels (c) and (e) are an artifact resulting from finite crystal-momentum resolution. }
 (g) Hole lifetime as a function of charge-carrier energies at 80, 298, 450, and 600~K. 
      The isotropic average at a given energy is shown by dashed lines, whereas 
      the shaded areas indicate energy windows of 50~meV, corresponding to the thermal energy $k_{\rm B}$T at 600~K for undoped (U)
      and doped (D) SnSe. 
      (h) Temperature-dependent renormalization of the effective mass 
      along the three crystal axes. The bare  DFT effective masses $m_{\rm b}(T)$
are shown by dashed lines.
 }
\label{fig:spec}
\end{figure*}

The Eliashberg function $\alpha^2F(\omega)$ for doped SnSe (here and below we 
consider a $p$-type doping concentration of $4\cdot10^{19}$~cm$^{-3}$), shown in Fig.~\ref{fig:phon}~(d)
reflects the weighted phonon density of states, 
to which individual phonons contribute according to their 
electron-phonon coupling strength \cite{grimvall1981electron}. 
$\alpha^2F(\omega)$ exhibits a sharp peak at 20~meV, 
which arises from the $B_g$ mode marked by red dots in 
Fig.~\ref{fig:phon}~(a). 
This feature accounts for about 50\% of the 
total electron-phonon coupling strength 
$\lambda = 2 \int_0^\infty d\omega \alpha^2F(\omega)/\omega$, 
suggesting that the coupling between electrons and phonons 
stems primarily from this mode. 
This scenario is reminiscent of highly-doped oxides, such as, e.g.,  
TiO$_2$ and EuO, where the formation of distinctive polaronic satellites 
in the angle-resolved spectral function \cite{Moser2013prl,Moser2015} 
has been attributed to the coupling to a single polar phonon \cite{Verdi2017,EuO,Caruso/PRB/2018}.

We now proceed to investigate the temperature dependence of the electronic structure. 
In Fig.~\ref{fig:spec}~(a)-(f), we show the 
angle-resolved spectral function due to the electron-phonon 
interaction in the diagonal approximation,
\begin{equation}\label{eq:Adiag}
 A_{n{\bf k}}(\omega) = -\frac{1}{\pi} \frac{ {\rm Im} \Sigma_{n{\bf k}}(\omega) }
 { [ \hbar\omega - \epsilon_{n{\bf k}} -  {\rm Re} \Sigma_{n{\bf k}}(\omega)]^2
 + [ {\rm Im} \Sigma_{n{\bf k}}(\omega)]^2} 
\end{equation}
at 80 and 600~K { for undoped SnSe}.
Numerical results for intermediate temperatures 
(298 and 450~K) are reported in the Supplemental Material \cite{sup}.
We consider paths parallel to the three crystallographic directions, 
passing through the valence-band maximum (VBM) at $(0,0,0.26)$~\AA$^{-1}$.
The single-particle energies obtained from DFT band-structure
calculations (light blue) are included for comparison.  
All energies are relative to the VBM. 

As compared to Eq.~\eqref{eq:Adiag}, 
the cumulant expansion approach provides a more suitable 
formalisms to describe photoemission satellites \cite{Gumhalter/2016/PRB}. 
At variance with highly-doped oxides, however, which exhibit  
signatures of low-energy plasmonic and polaronic satellites 
in ARPES \cite{Moser2013prl,Baumberger2016,SnO2, Verdi2017,Caruso/PRB/2018}, 
SnSe does not reveal such effects, thus, validating a description of 
its spectral properties within the Fan-Migdal approximation.

The comparison of the spectral function obtained from Eqs.~\eqref{eq:sigma}-\eqref{eq:Adiag} 
to the DFT band structure, where the electron-phonon interaction is neglected,
reveals a pronounced temperature-dependent renormalization of 
the electronic structure. 
In the low-temperature regime (80~K), the quasiparticle peaks 
are broadened by the interaction with phonons, reflecting the 
finite lifetimes of photo-excited holes in the valence band.
However, the spectrum exhibits only a small renormalization of 
the quasiparticle energies, as indicated by their 
overlap with the DFT bands.

The increase of temperature to 600~K is accompanied by a significant 
enhancement of the electron linewidths and of 
the quasiparticle energy renormalization. 
The linewidth is more pronounced for binding energies larger than 0.4~eV. 
This behaviour, which is in line with recent ARPES experiments performed at 
room temperature, is dictated by the onset of the $4p$-Se bands, 
which enlarges the phase-space available for the decay of photo-excited holes. 
At these binding energies, we obtain electron linewidths ranging between 0.2 and 0.4~eV. 
As they become comparable to the energy separation 
between different valence states, the identification of 
distinct quasiparticle bands is obstructed by the large peak 
broadening, and the spectral features merge together into a broad background. 
These results suggest a breakdown of the quasiparticle picture
at high temperatures, whereby a transition to a strong 
electron-phonon coupling regime is induced by the thermal 
excitation of soft polar modes. 

To further investigate the influence of temperature on the electronic properties,  
we evaluate the effective-mass renormalization due to the combined effect of  
(i) the electron-phonon interaction and 
(ii) the anisotropic thermal expansion of the crystal lattice \cite{th-exp/prb/2016}.
To account for these effects, we obtain  the effective mass from \cite{grimvall1981electron}:
\begin{align}
\label{eq:meff-renorm}
m^*(T)= m^*_{\rm b} (T) [1 + \lambda_{n{\bf k}} (T)].
\end{align}
In this expression, $m^*_{{\rm b}}$ is the bare effective mass
extracted from the DFT band structure. Its temperature dependence stems 
from the thermal expansion of the lattice induced by anharmonic effects, 
which are accounted for by employing at each temperature 
the experimental crystal structure \cite{th-exp/prb/2016}. 
{ The necessity of accounting for lattice anharmonicities in SnSe has
previously been suggested in Ref.~\cite{Verstraete/2016}, 
whereby the use of temperature-dependent crystal structures has proven 
important to obtain accurate Seebeck coefficients.}
The coupling to phonons is included via the electron-phonon coupling strength 
$\lambda_{n{\bf k}} = - \hbar^{-1} \partial 
\Sigma_{n{\bf k}} (\omega)/\partial\omega |_{\varepsilon_{\rm F}}$ 
with $\varepsilon_{\rm F}$ being the Fermi energy, {and $n,{\bf k}$ 
denoting the band and momentum index corresponding to the valence band top. }

The temperature dependence of $m_{{\rm b}}^*$ and 
$m^*$ relative to $m_{\rm b}^*(T=0)$
is illustrated in Fig.~\ref{fig:spec}~(h).
In the low-temperature limit, 
our calculations yield $m_{{\rm b}}^*(0) = $ 0.69, 0.34, and 0.16~$m_{\rm e}$ along $a$, $b$, and $c$, respectively, 
where $m_{\rm e}$ is the electron mass, in good agreement with earlier calculations \cite{Kioupakis/QP/2015}. 
The thermal expansion of the lattice leads to a monotonic decrease of 
$m_{{\rm b}}^* (T)$ with increasing temperature 
for momenta along the reciprocal lattice vectors ${\bf k}_{\rm a}$ and ${\bf k}_{\rm b}$. 
The opposite trend observed along ${\bf k}_{\rm c}$ stems 
from the negative thermal expansion of the lattice 
in this direction \cite{th-exp/prb/2016}. 
The inclusion of the electron-phonon interaction 
systematically enhances the effective mass with temperature, 
owing to the increase of the 
phonon population via thermal excitation. 
While the effective-mass renormalization due to 
the lattice expansion and the coupling to phonons sums up along ${\bf k}_{\rm c}$, 
these effects partially cancel out in the ${\bf k}_{\rm a}$ and ${\bf k}_{\rm b}$ directions.
Remarkably, along ${\bf k}_{\rm b}$ the interplay of lattice anharmonicity 
and electron-phonon coupling leads to a distinctive 
maximum of the effective mass at around 160~K, a 
trend in good agreement with recent ARPES measurements \cite{Kanatzidis/2018}.
Overall, these results reveal a complex interplay of 
lattice anharmonicity and electron-phonon interactions which underpins
a highly anisotropic temperature-dependent modulation 
of the effective mass. 
{Lattice anharmonicities and finite temperature effects 
may further influence the electron-phonon interaction via 
the change of the phonon dispersion for the thermally 
expanded structure. While finite temperature calculations
have been addressed in the past \cite{Hellman2011,Hellman2013}, accounting for such effects 
in {\it ab-initio} studies of the electron-phonon interaction 
is still a challenge.  }

\begin{figure}[t] 
  \includegraphics[width=0.48\textwidth]{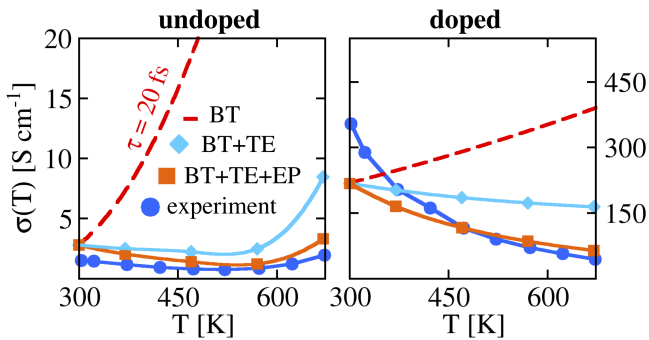} 
  \caption{Electrical conductivity along the crystallographic axis 
    $a$ for undoped (left) and doped (right) SnSe
    obtained using the Boltmann transport
    equation (BT) with a constant
    relaxation time $\tau = 20$~fs;
    BT with thermal expansion of the lattice (BT+TE); 
    and combining BT with thermal expansion and temperature-dependent $\tau$ due to electron-phonon
    coupling (BT+TE+EP).
    Experimental data (circles) are from Ref.~\cite{doping/science/2015}.}\label{fig:cond}
\end{figure}

The hole lifetimes, obtained as $\tau_{n{\bf k}} = \hbar/2{\rm Im}\Sigma_{n{\bf k}}$, 
and illustrated in Fig.~\ref{fig:spec}~(g) as a function of carrier energy for 
temperatures between 80 and 600~K,  further provide a quantitative estimation of 
the characteristic timescales for the relaxation of excited holes, 
which is critical  to infer the temperature-dependence of the transport properties.
The shaded region indicates the energy window of thermally excited 
carriers which take part in transport phenomena at 600~K for undoped and {doped} SnSe. 
Close to the Fermi energy, we obtain lifetimes of the order of 45-50~fs at 80~K, 
which drop below 10~fs at 600~K. 
The temperature-induced suppression of carrier dynamics 
is much more pronounced than in non-polar semiconductors 
and insulators, as e.g., diamond \cite{Logothetidis1992,Giustino2010prl}
and silicon \cite{Lautenschlager1986,Ponce2015}.
For binding energies larger than 0.4~eV, Fig.~\ref{fig:spec}~(g) 
illustrates a further suppression of the relaxation time, down to 8~fs 
(1~fs) at 80~K (600~K). 

Finally, to investigate how temperature affects the transport properties,
we evaluate the electrical conductivity 
$\sigma$ via the linearized Boltzmann equation in the relaxation time 
approximation \cite{electrontrans/PRB2003}:
\begin{align}\label{eq:cond}
{\boldsymbol{\sigma}} = 
\frac{e^2}{m_{\rm e}^2} \sum_{ n \bf k} \left( - \left. \frac{\partial f}{\partial \epsilon} \right\vert_{\epsilon_{n{\bf k}}} \right) {\bf p}_{n{\bf k}}{\bf p}_{n{\bf k}} \, \tau_{n{\bf k}}~.
\end{align}
Here, ${\bf p}_{n{\bf k}} = -i\hbar \langle \psi_{n{\bf k}}| \nabla | \psi_{n{\bf k}} \rangle $ 
is the momentum matrix element of the electronic state with band 
index $n$ and wave vector ${\bf k}$, 
$f$ the Fermi-Dirac distribution function, 
and $\tau_{n {\bf k}}$ the carrier scattering time.
In the following, we approximate $\tau_{n {\bf k}}$ by the isotropic 
average of the relaxation time at the Fermi surface. 
Calculations based on Eq.~\eqref{eq:cond} are 
performed for undoped and doped SnSe, using the {\tt exciting} 
code \cite{Gulans2014} (all computational details are reported in the 
Supplemental Material \cite{sup}) \footnote{Input and output files 
are available online at {\tt http://dx.doi.org/10.17172/NOMAD/2018.10.10-1}}, 
and compared to 
measurements from Ref.~\cite{doping/science/2015}. 
For the undoped SnSe, we assume a hole
concentration of $4.5 \cdot 10^{17}\,{\textrm{cm}}^{-3}$ {to account for 
the additional carriers induced by the formation of Sn vacancies \cite{Verstraete/2016}.}
The results obtained for the crystallographic axis ${\bf a}$
are shown in Fig.~\ref{fig:cond}. 

If the temperature dependence of the relaxation time is neglected  
(setting  $\tau= 20\,\textrm{fs}$) in Eq.~\eqref{eq:cond}, 
the standard expression for the conductivity  
Boltzmann transport (BT) of both undoped and 
hole-doped SnSe increases with temperature 
owing to the enhanced population of thermally-excited holes. 
This behavior, which is in striking contrast with experiment \cite{doping/science/2015}, 
is ameliorated by taking into account the thermal lattice expansion 
at each temperature (BT+TE, squares in Fig.~\ref{fig:cond}).
However, a good quantitative agreement with the experimental conductivity
is only recovered by simultaneously accounting for lattice anharmonicities 
and the temperature-dependent renormalization of the relaxation 
time due to the electron-phonon interaction (BT+TE+EP).
The residual difference between theory and experiment is ascribed to  
uncertainties in the determination of the hole concentrations 
and to additional coupling mechanisms, such as, 
e.g., impurity, electron-electron, and electron-plasmon 
scattering \cite{PhysRevB.94.115208}, neglected in our calculations.
Overall, our results indicate that, beside influencing pervasively 
the spectral properties, the thermal enhancement of the 
polar-phonon population is also responsible for the 
temperature-induced suppression of hole transport. 

In conclusion, we have investigated the influence of temperature on the 
electronic and transport properties of SnSe by accounting for the 
coupling between electrons, phonons, and lattice anharmonicities 
within a first-principles many-body framework.
At finite temperature the electronic structure exhibits 
thermally-enhanced electron linewidths, a suppression 
of hole relaxation times, and a highly-anisotropic 
renormalization of the hole effective masses.
These phenomena are attributed to the coexistence of 
soft polar phonons and large Born effective charges, 
which induce a strong increase of the electron-phonon 
interaction between 0 and 600~K.
Accounting simultaneously for the thermal expansion of the lattice 
and electron-phonon interaction is key to explain the temperature 
dependence of recent experimental data for the electrical conductivity. 
Overall, a sizeable thermal enhancement of the electron-phonon 
interaction may constitute a general feature of compounds
characterized by low-energy phonons with large Born effective 
charges such as, e.g., other Pb- and Sn-chalcogenides \cite{Zhang2009,Tian2012,Guo2015}. 
This study reveals the crucial importance of accounting for 
these processes to accurately describe transport phenomena in 
high-performance thermoelectrics.

\acknowledgments
Discussions with Dino Novko are gratefully acknowledged. 
MT acknowledges funding from the Elsa-Neumann Scholarship of Berlin.


%

\end{document}